\documentclass[aps,10pt,pra,twocolumn,floatfix,floats,showpacs,superscriptaddress,raggedbottom]{revtex4-1}
\usepackage{graphicx,latexsym}
\usepackage{dcolumn}
\usepackage{amsmath}
\usepackage{amssymb,bm}
\usepackage{braket}
\usepackage{natbib}

\usepackage[normalem]{ulem}

\def\mb#1{\mbox{\boldmath$#1$}}

\def\sec#1{Sec.\ \ref{#1}}
\def\eq#1{Eq.\ (\ref{#1})}
\def\fig#1{Fig.\ \ref{#1}}

\begin{document}
%-----------------------------------------------------------------

\title{Thermoelectric inversion in a resonant quantum dot-cavity system\\ in the steady-state regime}

\author{Nzar Rauf Abdullah}
\email{nzar.r.abdullah@gmail.com}
\affiliation{Physics Department, College of Science, 
             University of Sulaimani, Kurdistan Region, Iraq}
\affiliation{Komar Research Center, Komar University of Science and Technology, Sulaimani City, Iraq}

\author{Chi-Shung Tang}
\affiliation{Department of Mechanical Engineering,
  National United University, 1, Lienda, Miaoli 36003, Taiwan}

\author{Andrei Manolescu}
\affiliation{Reykjavik University, School of Science and Engineering,
              Menntavegur 1, IS-101 Reykjavik, Iceland}

\author{Vidar Gudmundsson}
\email{vidar@hi.is}
 \affiliation{Science Institute, University of Iceland,
        Dunhaga 3, IS-107 Reykjavik, Iceland}

%
%----------------------------------------------------------------

\begin{abstract}
The thermoelectric effect in a quantum dot system connected to two electron reservoirs
in the presence of a photon cavity is investigated using a quantum master equation in 
the steady-state regime.
If a quantized photon field is applied to the quantum dot system, an extra channel, 
the photon replica states, are formed leading to a generation of a photon-induced 
thermoelectric current. 
We observe that the photon replica states contribute to the transport 
irrespective of the direction of the thermal gradient. 
In the off-resonance regime, when the photon energy is smaller than the energy 
difference between the lowest states of the quantum dot system, a current plateau is seen for  
strong electron-photon coupling.
In the resonant regime, an inversion of thermoelectric current emerges due to 
the Rabi-splitting. Therefore, the photon field can change both the magnitude 
and the sign of the thermoelectric current induced by the temperature gradient in 
the absence of a voltage bias between the leads.

\end{abstract}

%\pacs{73.23.-b, 73.21.Hb, 75.47.-m, 85.35.Ds}

% 73.23.-b: Electronic transport in mesoscopic systems
% 73.21.Hb: Quantum wires
% 75.47.-m Magnetotransport phenomena; materials for magnetotransport
% 85.35.Ds: Quantum interference devices

\maketitle

%
%------------------------------------------------------
%

\section{Introduction}\label{Sec:Introduction}

Thermoelectric transport through nanoscale systems has been studied experimentally~\cite{0957-4484-26-3-032001,PhysRevLett.95.176602}
and theoretically~\cite{refId0, PhysRevB.65.115332, PICHARD20161039} with the aim to control heat flow and harvesting thermal energy.
Special interest has been on the thermoelectric properties of quantum dots (QD) in the Coulomb blockade (CB) regime 
both in weakly and strongly coupled QD-devices~\cite{PhysRevLett.86.280, PhysRevB.46.9667,PhysRevB.90.115313}.
Several approaches have been used to increase the efficiency of such devices. 
The efficiency of the QD system has been studied considering either the Coulomb interaction in two- or 
multi-level QD using non-equilibrium Green's function (NEGF) methods~\cite{PhysRevB.81.245323} or the electron-phonon 
contribution~\cite{0953-8984-22-35-355304}. A general formula modeling the heat current in a lead-QD-lead system by NEGF has been proposed 
and it has been shown that the heat current could be very high in the Coulomb blockade regime in which the thermoelectric current 
is very small due to the Coulomb blockade effect~\cite{PhysRevB.79.161309}.

Besides the conventional thermoelectric nano-structures, spintronic devices have been employed to enhance 
and control the efficiency of the thermoelectric transport using the spin degree of freedom in addition 
to charges~\cite{PhysRevB.95.165439, Hauptmann2008}. To build a spintronic nanoscale system, spin 
polarized electrons has to be considered. One approach has been to take into account the Rashba-spin orbit coupling 
in the dot system \cite{Bathen2017,Abdullah2017} 
or assuming ferromagnetic lead-based spintronic devices~\cite{YAN2016277}. 
In both cases, the spin effects can cause increase in the figure of merit and thermal conductance which 
are important in controlling the performance of nano-devices.

Another technique to control thermoelectric efficiency is to use a photon field. 
The thermoelectric transport between two bodies, mediated by electromagnetic fluctuations, can be controlled with 
an intermediate quantum circuit leading to the device concept of a mesoscopic photon heat transistor~\cite{PhysRevLett.100.155902}.
The proposed thermal quantum transistor could be used to develop devices such as a thermal modulator and
a thermal amplifier in nanosystems~\cite{PhysRevLett.116.200601}. Furthermore, it has been shown that heat can be conducted 
by a photon field at a very low temperature when the phonons are frozen out~\cite{Meschke2006} and 
the photon field can change both the magnitude and the sign of the electrical bias voltage
induced by the temperature gradient~\cite{0953-8984-24-14-145301} which plays the role of a thermal amplifier.

Based on aforementioned investigations, we study thermoelectric transport through a QD system coupled to a photon cavity, 
where the QD system is either in a resonance or off-resonance with the photon field. 
A Markovian version of a Nakajima-Zwanzig generalized master equation is utilized to investigate the transport 
properties of the system~\cite{Nakajima20.948,Zwanzing.33.1338,JONSSON201781}.
In previous publications, we have reported that the transient thermoelectric~\cite{Nzar_ACS2016} and heat~\cite{ABDULLAH2018199} 
currents can be modulated
using a cavity photon field with even a single photon~\cite{ABDULLAH2018223}. 
The influences of the photon polarization and the electron-photon coupling strength on the 
transient thermospin current and spin-dependent heat current in different systems have been demonstrated~\cite{ABDULLAH2018102,
0953-8984-30-14-145303,ABDULLAH2018}. 

In this work, we investigate the thermoelectric current through a QD system in the steady-state regime. 
We show the effects of a photon field such as polarization and electron-photon coupling strength on the thermoelectric current. 
Thermoelectric current oscillations ``peak'' is observed due to the photon-assisted tunneling processes. In addition, 
thermoelectric current plateau in the off-resonance regime and current inversion in the resonant regime are found.

The outline of the paper is as follows. We describe the model system in~\sec{Sec:Model}.
Results are discussed for the model in~\sec{Sec:Results}. Finally, we draw our conclusion in~\sec{Sec:Conclusion}.

\section{Model and Transport formalism}\label{Sec:Model}

\begin{figure}[htb]
 \includegraphics[width=0.45\textwidth]{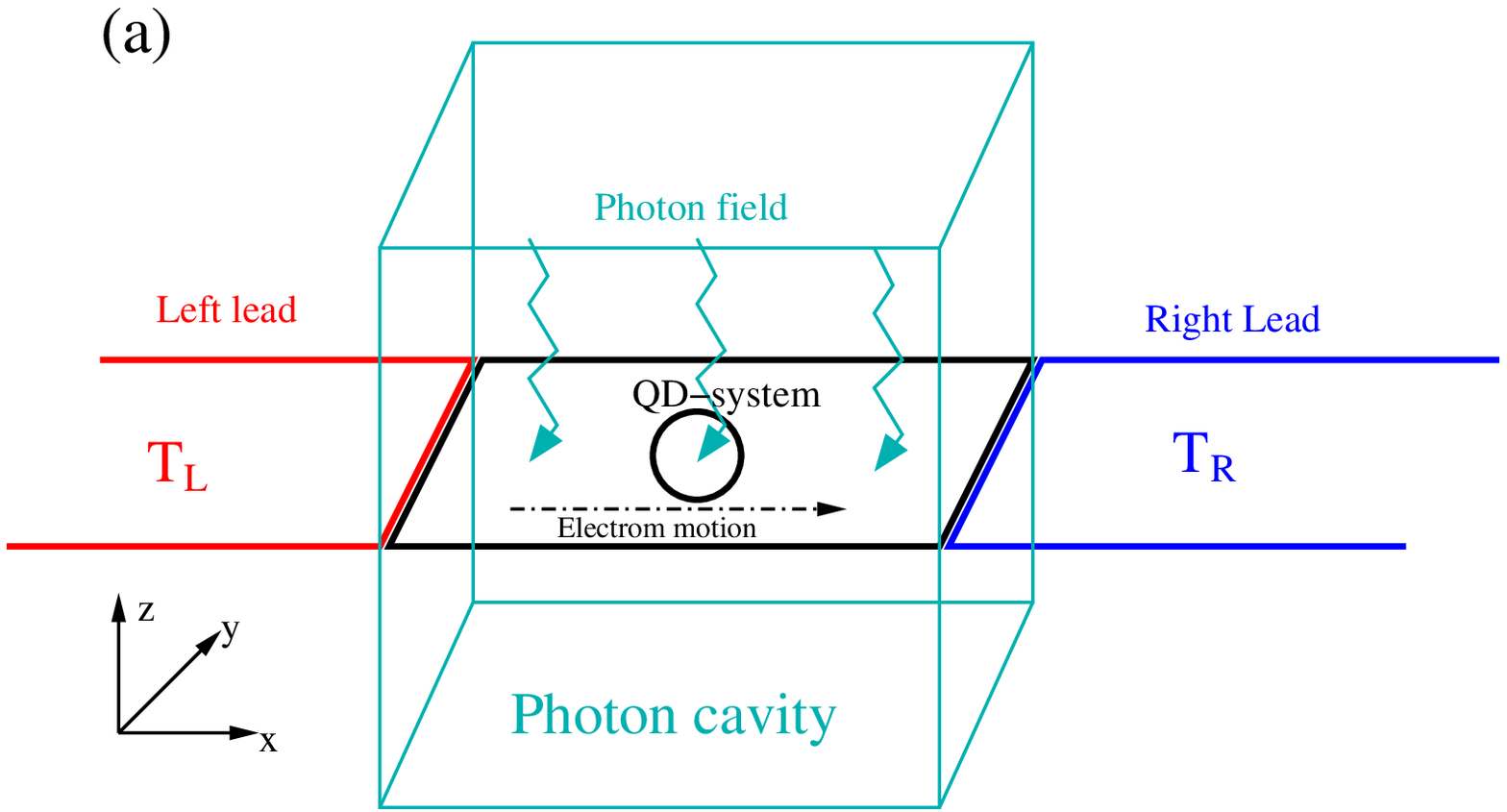}
 \includegraphics[width=0.4\textwidth]{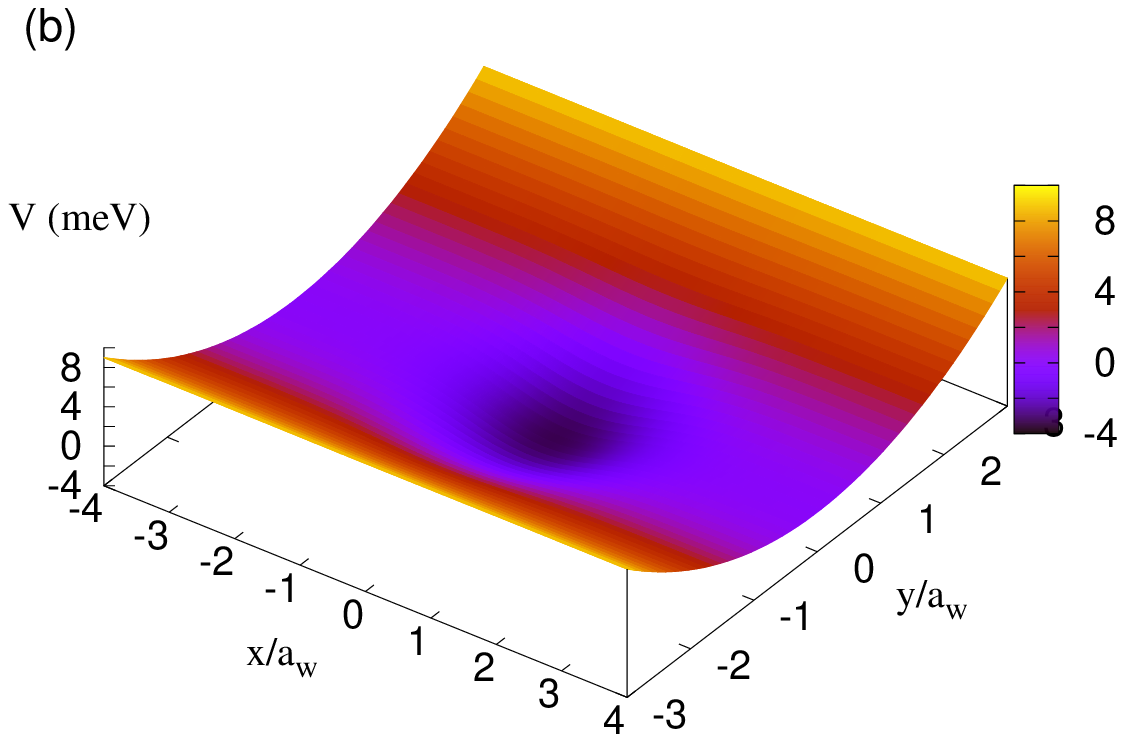}
\caption{ (a) Schematic diagram showing the QD-system (black) coupled
 to the leads, where the temperature of the left lead ($T_L$) (red) is higher than the temperature of the 
 right lead ($T_R$) (blue). 
 The cyan zigzag display the photon field in the cavity (cyan rectangle). (b) The potential $V_r(\mathbf{r})$ defining the central QD system
 that will be coupled diametrically to the semi-infinite left and right leads
 in the $x$-direction.}
\label{fig01}
\end{figure}

In this section, the model and the theoretical formalism used to calculate 
the thermal properties of the system are presented.
We assume a quantum dot embedded in a two dimensional quantum wire, and the QD-system 
is coupled to two semi-infinite leads from both ends as is shown in \fig{fig01}(a). The cyan zigzag arrows indicate the 
photon field inside the photon cavity (cyan rectangle) coupled to the QD-system.  The temperature of the left lead ($\rm T_{L}$) 
(red color) is considered to be higher than that of the right lead ($\rm T_{R}$) (blue color). 
The contact regions couple the QD-system and the leads.
Figure \ref{fig01}(b) shows the potential of the quantum dot embedded in the two-dimensional quantum wire in the $xy$-plane.
The electrons are mainly transferred through the QD system in the $x$-direction by the thermal bias.

The potential used to define the QD-system can be represented as 
\begin{align}
 V_{\rm QD}(x,y) & =  \Big[ \frac{1}{2} m^* \Omega^2_0 y^2 + eV_{\rm p}
            + V_0 \exp(-\beta_x^2 x^2 - \beta_y^2 y^2) \Big] \nonumber \\
            & \times \theta \Big(\frac{L_x}{2} - |x| \Big) ,
\end{align}
where $\Omega_0$ is the electron confinement frequency due to the lateral parabolic potential, 
$m^*$ refers to the effective mass of the electrons,
$V_{\rm p}$ indicates the plunger-gate voltage that shifts the energy states of the QD-system 
with respect to the chemical potential of the leads, and $\theta$ is the Heaviside unit step function with the 
length of the quantum wire $L_x = 150$~nm. 
Here, we assume $V_0 = -3.3$ meV and $\beta_x = \beta_y = 0.03$~nm$^{-1}$ determining the diameter of the QD.

The QD-system is hard-wall confined in the $x$-direction and parabolocally confined in
the $y$-direction. The Hamiltonian for the two-dimensional system can be described by~\cite{2040-8986-17-1-015201,PhysRevE.86.046701,Nzar_2016_JPCM} 
\begin{align}\label{H_S}
 \hat{H}_{S}&=\int d^2r\;\hat{\psi}^{\dagger}(\mathbf{r}) \left[\frac{1}{2m^{*}}\left(\frac{\hbar}{i}\nabla+
              \frac{e}{c}\left[\mathbf{A}_{\rm B}(\mathbf{r}) +\hat{\mathbf{A}}_{\gamma}(\mathbf{r})\right] \right)^2 \right. \nonumber \\
            & \left. + V_{\rm QD}(\mathbf{r}) \right] \hat{\psi}(\mathbf{r}) + H_Z +
              \hat{H}_{ee} +\hbar \omega_{\gamma} \hat{a}^{\dagger}\hat{a}.
\end{align}
The electron field operator is $\hat{\psi}$, the magnetic vector potential is $\mathbf{A}_{\rm B}(\mathbf{r})= -By \hat{\mb{x}}$ defined in the Landau gauge, and
$\hat{\mathbf{A}}_{\gamma}$ is the vector potential of the photon cavity given by 
\begin{equation}\label{A}
 \hat{\mathbf{A}}_{\gamma}(\mathbf{r})=A(\hat{a}+\hat{a}^{\dagger}) \mathbf{e}
\end{equation}
with $A$ the amplitude of the photon cavity field determined by the electron-photon coupling strength $g_{\gamma}=eA a_w \Omega_w/c$,
$\mathbf{e} = \mathbf{e}_x$ for parallel polarized photon field ($\mathrm{TE}_{011}$) or $\mathbf{e} = \mathbf{e}_y$ for 
perpendicular polarized photon field ($\mathrm{TE}_{101}$), and $\Omega_w$ the effective confinement frequency
determined by the external static magnetic field $\mathbf{A}_B$ and the bare confinement energy $\hbar\Omega_0=2.0$ meV.
The photon creation and annihilation operators are represented by $\hat{a}^{\dagger}$ and $\hat{a}$, respectively.

The second part of \eq{H_S} is the Zeeman Hamiltonian defining the interaction between the magnetic moment of an electron and 
the weak external magnetic field, $H_Z = \pm g^{*}\mu_B B/2$, with $\mu_B$ the Bohr magneton and $g^{*} = -0.44$ the 
effective g-factor for GaAs. The role of this magnetic field is to lift the spin degeneracy, which otherwise may create numerical difficulties.
The external magnetic field and the parabolic confinement in the $y$-direction define a characteristic length scale,
the effective confinement or magnetic length $a_w=(\hbar/(m^*\Omega_w))^{1/2}$.
In addition, the third part of \eq{H_S} ($\hat{H}_{ee}$)  
indicates the Coulomb interaction in the QD-system \cite{Nzar.25.465302},
and the last part is the quantized photon field, with $\hbar\omega_{\gamma}$ as the photon energy.

To investigate the transport properties of the system in the steady-state, 
we use a projection formalism built on the density operator \cite{Zwanzing.33.1338,Nakajima20.948}. 
The density operator of the total system before coupling to the leads is defined as the tensor product of
the individual density operators $\hat{\rho}(t < t_0) = \hat{\rho}_\mathrm{L} \hat{\rho}_\mathrm{R} \hat{\rho}_\mathrm{S}(t < t_0)$
where $\hat{\rho}_\mathrm{L}$ and $\hat{\rho}_\mathrm{R}$ indicate the density operator of the left (L)  and the right (R) leads, respectively.
After coupling the QD-system to the leads, one can find the reduced density operator 
$\hat{\rho}_\mathrm{S}$ that introduces the state of the electrons in the QD-system under the 
effect of the leads as
\begin{equation}
 \hat{\rho}_\mathrm{S} = {\rm Tr_\mathrm{ L,R}}[\hat{\rho}(t)], 
\end{equation}
where the trace is over the Fock space of the leads. We derive the equation of motion for the 
reduced density operator as a non-Markovian integrodifferential equation with a kernel evaluated
up to second order in the system-lead coupling \cite{Vidar61.305}. As we are interested in the
long-time evolution and the steady state of the system we further transform with equation into 
a corresponding Markovian equation for the reduced density operator of the QD-system \cite{JONSSON201781}
\begin{align}
      \partial_t\hat{\rho}_\mathrm{S}(t) = &-\frac{i}{\hbar}[\hat{H}_\mathrm{S},\hat{\rho}_\mathrm{S}(t)]
      -\left\{\Lambda^L[\hat{\rho}_\mathrm{S} ;t]+\Lambda^R[\hat{\rho}_\mathrm{S} ;t]\right\}\nonumber\\
      &-\frac{\bar{\kappa}}{2\hbar}(\bar{n}_\textrm{R}+1)\left\{2\alpha\hat{\rho}_\mathrm{S}\alpha^\dagger - \alpha^\dagger \alpha\hat{\rho}_\mathrm{S} 
                                                                                 - \hat{\rho}_\mathrm{S}\alpha^\dagger \alpha\right\}\nonumber\\
       &-\frac{\bar{\kappa}}{2\hbar}(\bar{n}_\textrm{R})\left\{2\alpha^\dagger\hat{\rho}_\mathrm{S}\alpha  - \alpha\alpha^\dagger\hat{\rho}_\mathrm{S} 
                                                                                 - \hat{\rho}_\mathrm{S}\alpha\alpha^\dagger\right\}.
\label{NZ-eq}
\end{align}
Herein, $\Lambda^L$ and $\Lambda^R$ represent the ``dissipation'' processes caused by both leads, 
$\bar{\kappa} = 1.0 \times 10^{-5}$~meV is the photon decay constant, $\bar{n}_\textrm{R}$ indicates the mean photon number of the reservoir.
The second and the third line in Eq.\ \ref{NZ-eq} embody the photon dissipation of the cavity. 
$\alpha^\dagger$ ($\alpha$) stands for the original operator in the non-interacting photon number basis, $a^\dagger$ ($a$), transformed 
to the interacting electron photon basis using the rotating wave approximation~\cite{GUDMUNDSSON20181672}, where care has been taken 
in constructing a non-white noise spectrum appropriate for strong electron-photon coupling 
\cite{PhysRev.129.2342,PhysRevA.31.3761,PhysRevA.84.043832}.

The QD-system is coupled to the two leads that play the role of electron reservoirs 
obeying the Fermi-Dirac distribution
\begin{equation}\label{Eq_1}
 F_\mathrm{L,R} = \Big[ 1 + \exp{\big((E-\mu_\mathrm{L,R})/(k_{\rm B} T_\mathrm{L,R}) \big)} \Big]^{-1}.
\end{equation}
with $\mu_\mathrm{L}$ ($\mu_\mathrm{R}$) the chemical potential of the left (right) leads, respectively, and 
$T_{\rm L}$ and $T_{\rm R}$ are temperature of the left and right leads, respectively. 
The Fermi distribution of the leads is included in both dissipation terms $\Lambda^L$ and $\Lambda^R$~\cite{JONSSON201781}.

We consider the chemical potential of the leads to be equal ($\mu_\mathrm{L} = \mu_\mathrm{R}$) here and the temperature 
of the left lead to be higher than the temperature of the right lead. 
Therefore, the temperature gradient generates thermoelectric current through the QD-system coupled to the leads. 
The thermoelectric current from the left lead into the QD-system, $I_{\rm L}$,
and the thermoelectric current from it into the right lead, $I_{\rm R}$, can 
be introduced as 
\begin{equation}
 I_\mathrm{L,R} = {\rm Tr}_\mathrm{S} \Big( \Lambda^{\rm L,R}[\hat{\rho}_\mathrm{S};t] Q \Big).
\end{equation}
The charge operator of the QD-system is $Q = -e \sum_i d_i^\dagger d_i$ with $\hat{d}^\dagger (\hat{d})$
the electron creation (annihilation) operator of the central system, respectively.

\section{Results}\label{Sec:Results}

The QD-system and the lead are made of a GaAs material with the effective mass $m^* = 0.067 {\rm m_e}$ 
and relative dielectric constant $\kappa = 12.4$.
The effective confinement frequency is $\Omega_w = \sqrt{\Omega^2_0 + \omega^2_c}$ with $\Omega_0$ being electron confinement frequency
due to the lateral parabolic potential and $\omega_c$ the cyclotron frequency due to external magnetic field. 
The electron confinement energy in both the QD-system and the leads is considered to be $\hbar \Omega_0 = \hbar\Omega_{\rm L,R} = 2.0$~meV, 
and the cyclotron energy is $\hbar\omega_c = 0.172$~meV at the weak external magnetic field $B = 0.1$~T applied to the total system
leading to $a_w=23.8$ nm.

\begin{figure}[htb]
 \includegraphics[width=0.30\textwidth]{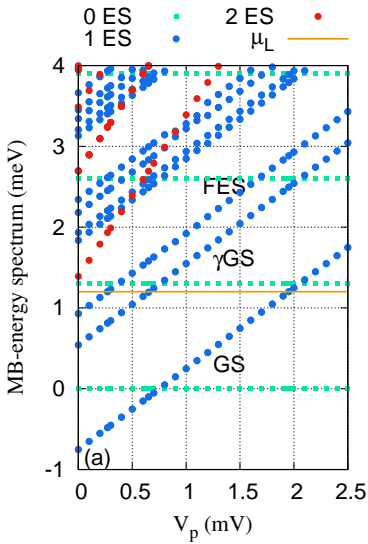}
\caption{MB-Energy spectrum ($\rm E_{\rm \mu}$) of the quantum dot system versus the gate voltage ($\rm V_p$), where 
        0ES (green squares) are zero-electron states, 1ES (blue circles) are one-electron states, 
        and 2ES (red circles) are two-electron states.
         The golden line is the chemical potential of the leads where $\mu_L = \mu_R = \mu = 1.2~{\rm meV}$.
         GS indicates the one-electron ground-state, $\gamma$GS is the one-photon replica of the one-electron ground-state, and 
        FES is the one-electron first-excited state.
        The photon energy $\hbar \omega_{\gamma} = 1.31$~meV, the electron-photon coupling strength is $g_{\gamma} = 0.05$~meV and
        and the photon field is linearly polarized in the $x$-direction. 
         The magnetic field is $B = 0.1~{\rm T}$, and $\hbar \Omega_0 = 2.0~{\rm meV}$.}
\label{fig02}
\end{figure}

Figure \ref{fig02} shows the Many-Body (MB) energy spectrum versus the plunger-gate voltage $V_{\rm p}$ 
for the QD-system coupled to the cavity. The golden horizontal line indicates the chemical potential of the leads 
$\mu_{\rm L} = \mu_{\rm R} = 1.2$~meV. It is clearly seen that the ground-state (GS) at $V_{\rm p} = 1.95$~mV 
and the first-excited state (FES) at $V_{\rm p} = 0.271$~mV are touching (reaching) the chemical potential of the leads.
Therefore, one can expect that these two states are responsible for the electron transport in the selected range of the gate voltage 
in the case of no photon cavity.
The photon energy is assumed to be $\hbar \omega_{\gamma} = 1.31$~meV 
which is smaller than the energy spacing between the GS and the FES at $g_{\gamma} = 0.05$~meV. 
Under these conditions, the QD-system is not in a resonance with the photon field. 
In addition to the two major states, the GS and the FES, there appear photon replica states. 
For instance, the lowest photon replica of the ground-state ($\gamma$GS) appearing in the energy spectrum
is in resonance with the chemical potential of the leads at $V_{\rm p} = 0.65$~meV. We note that 
the energy spectrum for the $x$- and the $y$-polarized photon field is almost the same here.

\begin{figure}[htb]
  \includegraphics[width=0.5\textwidth,angle=0,bb=70 104 410 210]{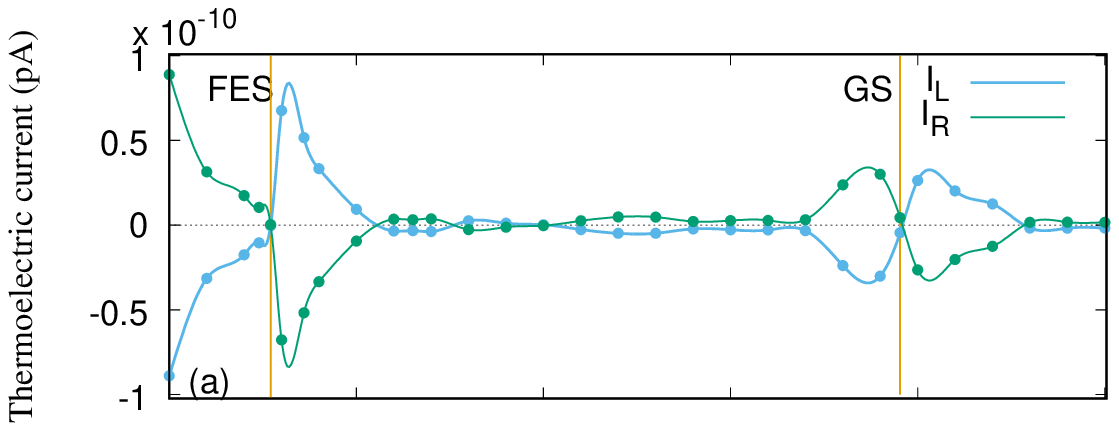}\\
  \includegraphics[width=0.5\textwidth,angle=0,bb=50 55 410 204]{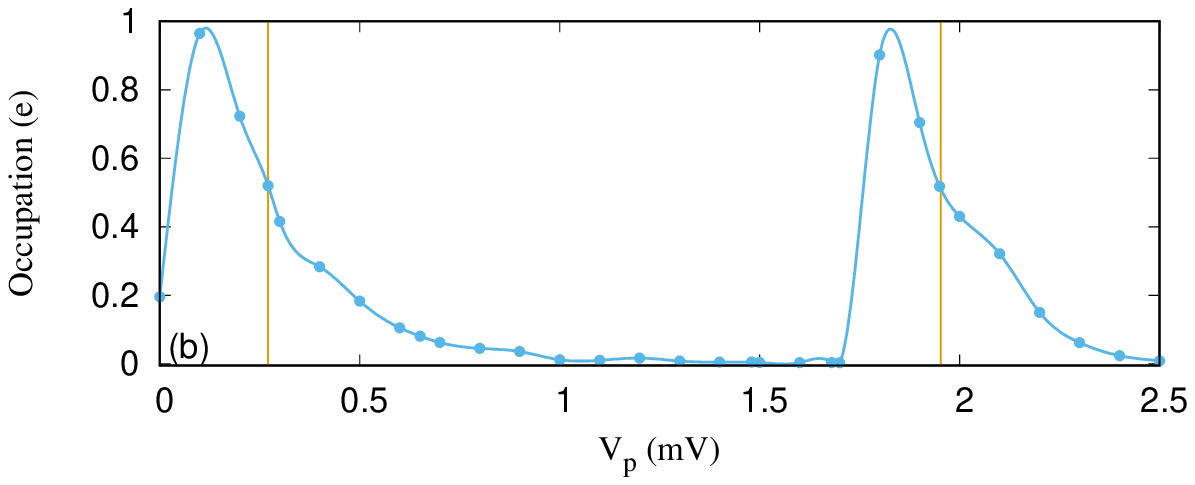}
 \caption{Thermoelectric current from the left lead to the QD-system ($I_L$) and the thermoelectric current 
               from the QD-system to the right lead ($I_R$) (a), and occupation (b) as functions of the gate voltage $\rm V_{p}$ for 
               the QD-system without the photon cavity.
               The temperature of the left (right) lead is fixed at ${\rm T_L} = 1.16$~K (${\rm T_R} = 0.58$~K) 
               implying thermal energy $0.1$~meV ($0.05$~meV), respectively.
               The chemical potential of the leads are fixed at $\mu_{\rm L} = \mu_{\rm R} = 1.2$~meV.
               The golden vertical lines show the resonance condition for the GS at ${\rm V_{p}} = 1.95$~mV, and 
               the FES at ${\rm V_{p}} = 0.271$~mV.
               The magnetic field is $B = 0.1~{\rm T}$, and
               $\hbar \Omega_0 = 2.0~{\rm meV}$.}
\label{fig03}
\end{figure}

To understand the properties of the thermoelectric current due to the temperature gradient
we start by considering the case of no photon cavity. 
In this case, the relevant states contributing to the transport are the original pure electron states, such as 
the GS and the FES. The left thermoelectric current, $I_{\rm L}$, into the QD-system 
and the right thermoelectric current, $I_{\rm R}$, out of it,
for these two states as a function of the gate voltage are presented in \fig{fig03}(a).
The left and the right thermoelectric currents are equal but with opposite sign indicating the onset
of a steady-state regime already at time just before $t = 1 \times 10^{8}$~ps, even though we follow 
the evolution to $t = 1 \times 10^{11}$~ps. 

The thermoelectric current emerges due to the difference between the two Fermi functions of the leads. 
The thermoelectric current is observed when the Fermi functions of the leads have 
the same chemical potential but different widths. It can be described as the follows: 
Thermoelectric current is zero in two situations. 
First, when the two Fermi functions of the leads or their occupations (see \fig{fig03}b) are equal to $0.5$ (half filling). 
Second, when both Fermi functions or occupations are $0$ or $1$ (integer filling)~\cite{Tagani201336, PhysicaE.53.178}.
As a result, the thermoelectric current is approximately zero at $V_{\rm p} = 0.271$ and $1.95$~mV 
corresponding to half filling of the FES and the GS, respectively~\cite{Nzar_ACS2016}. 
The thermoelectric current is approximately zero at $V_{\rm p} = 1.8$ and $2.4$~mV for an integer filling or occupation 
of 0 and 1, around the GS, respectively.

We note that the electron or charge occupation of the system is large when the GS or the FES are in or close to 
resonance with the the chemical potential of the leads. Without cavity photons the charge almost exclusively resides
in the corresponding resonant states and is vanishingly small for ${\rm V_{p}}$ in the range between 1.0 and 1.7 mV.
This can be understood having in mind that the temperatures ${\rm T_L}$ and ${\rm T_R}$ are very low, the GS localized
in the quantum dot is very weakly coupled to the leads, and the electron density of states of the quasi 1D leads has 
a peak at the lowest subband bottom at 1.0 meV, while the GS is well below this value for this range of the ${\rm V_{p}}$.
The coupling to the leads depends on the spatial extension of the corresponding wavefunctions into the contact areas
of width $a_w$ at the ends of each subsystem, respectively. In addition, the coupling depends on the electron affinity
defined by $\exp(-|E_a-\epsilon(q)|/\Delta_E)$, where $E_a$ stands for the states of the original single-electron
basis for the central system, $\epsilon(q)$ the energy spectrum of a lead, and $\Delta_E = 0.5$ meV here 
\cite{Vidar11.113007}.
The electron occupation or charge cumulation in the central system will be strongly affected by the cavity photon
field as will be reported below.

Let us now consider the situation when a photon field is applied to the QD-system. 
In the off-resonant regime, when the photon energy is assumed to be  $\hbar\omega_{\gamma} = 1.31$~meV which is 
smaller than the energy spacing between the two lowest state of the QD-system (${\rm E_{FES}} - {\rm E_{GS}} = 1.682$~meV) for 
$g_{\gamma} = 0.05$~meV and $x$-polarized photon field.

\begin{figure}[htb]
  \includegraphics[width=0.5\textwidth,angle=0,bb=70 104 410 210]{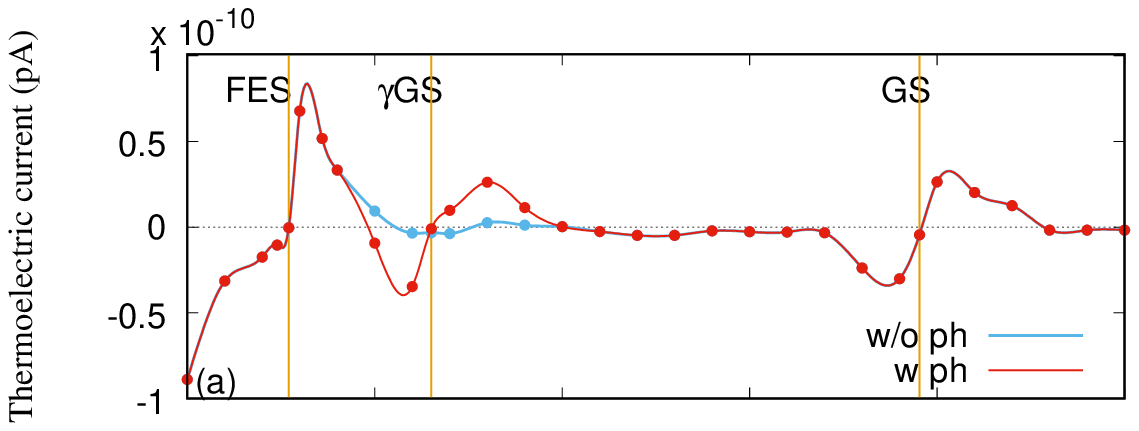}\\
  \includegraphics[width=0.5\textwidth,angle=0,bb=70 104 410 202]{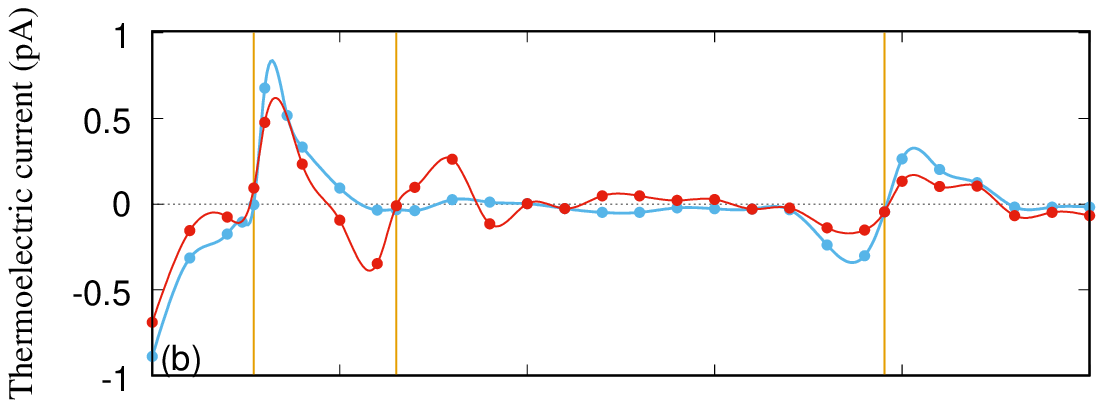}
  \includegraphics[width=0.5\textwidth,angle=0,bb=50 55 410 202]{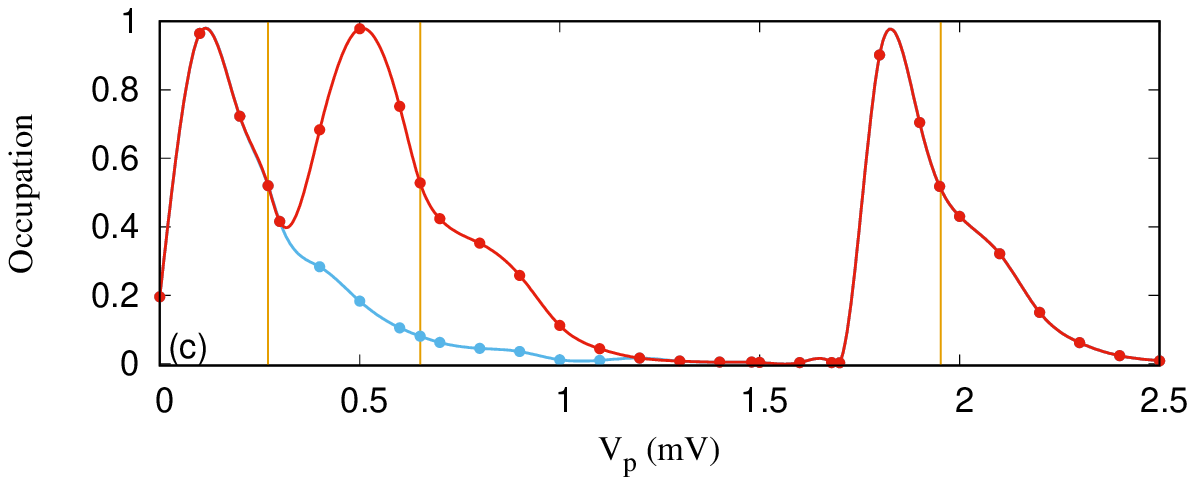}\\
 \caption{Thermoelectric current ($I_{\rm L}$) in the case of $\bar{n}_\textrm{R} = 0$ (a) and 
               $1$ (b), and electron occupation (c) for the QD-system without (w/o ph) (blue color) and with (w ph) (red color) photon field.  
               The photon energy is $\hbar\omega_{\gamma} = 1.31$~meV (off-resonance regime), $g_{\gamma} = 0.05$~meV, 
               and the photon field is polarized in the $x$-direction.
               The temperature of the left (right) lead is fixed at ${\rm T_L} = 1.16$~K (${\rm T_R} = 0.58$~K) 
               implying thermal energy $0.1$~meV ($0.05$~meV), respectively.
               The chemical potential of the leads are fixed at $\mu_{\rm L} = \mu_{\rm R} = 1.2$~meV.
               The golden vertical lines show the resonance condition for the GS at ${\rm V_{p}} = 1.95$~mV, 
               the $\gamma$GS at ${\rm V_{p}} = 0.65$~mV, and the FES at ${\rm V_{p}} = 0.271$~meV, respectively.
               The weak external magnetic field is $B = 0.1~{\rm T}$, and $\hbar \Omega_0 = 2.0~{\rm meV}$.}
\label{fig04}
\end{figure}

Figure \ref{fig04} demonstrates the left thermoelectric current ($I_{\rm L}$) for the off-resonant regime when
the mean photon number is $\bar{n}_\textrm{R} = 0$ (a) and $1$ (b). In addition, it's occupation 
versus the gate voltage is shown in \fig{fig04}(c). The occupation is almost the same for 
both cases of $\bar{n}_\textrm{R} = 0$, and $1$. Compared to the case of no photon field (blue color), 
extra current oscillation, from negative to positive, around the $\gamma$GS at $V_{\rm p} = 0.65$~mV  
is observed in the presence of the photon field for both $\bar{n}_\textrm{R} = 0$, and $1$. 
The additional current oscillation arises due to a photon-assisted tunneling (PAT)~\cite{PhysRevB.87.085427}.  
An additional ``peak'' in the occupation around $V_{\rm p} = 0.65$~mV shown in \fig{fig04}(c) is found corresponding 
to the extra current oscillation. 
The photon-assisted thermal transport has also been calculated for simple two level system using a 
Green function formalism~\cite{BAGHERITAGANI201386,TAGANI201336_PAT}.
We have not seen the extra thermal current peak in the transient regime~\cite{Nzar_ACS2016,ABDULLAH2018199}, 
however the photon-assisted charge current peak can be clearly seen in the transient regime~\cite{Nzar.25.465302,PhysicaE.64.254}.

\begin{figure}[htb]
  \includegraphics[width=0.25\textwidth]{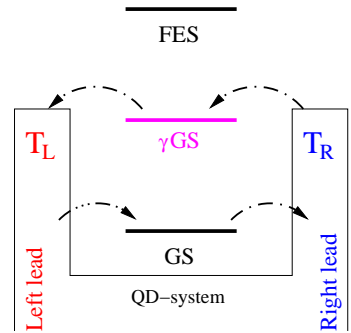}
 \caption{Schematic representation of photon activated
                         resonance energy levels and electron transition by
                         the photon-induced processes via $\gamma$GS}
\label{fig05}
\end{figure}

We should mention that the thermoelectric current is almost unchanged when $\bar{n}_\textrm{R} = 0$ and
a suppression of thermoelectric current around GS and FES for $\bar{n}_\textrm{R} = 1$ is recorded
due to the contribution of their photon replica states to the transport. 
The processes of current transport in the presence of the photon field is totally different here.
For example, the contributed ratio of the GS to the transport is approximately $90\%$ in the range
$V_{\rm p} = [1.8 - 2.2]$~mV where there is no photon field. But the GS is not longer the most active state that is responsible for 
the transport in the presence of the photon field. $\gamma$GS together with the GS contribute to the transport 
in this range  $V_{\rm p} = [1.8 - 2.2]$~mV and the mechanism of thermal transport is totally different for these 
two states. Thermoelectric current flows from the left lead to the right lead through the GS. Surprisingly, 
the direction of current through the $\gamma$GS is contrary going from right lead to the left lead as is shown in \fig{fig05} which is 
irrespective of the direction of the thermal gradient.
Therefore, the thermoelectric current is suppressed. 
The reversed transport via the $\gamma$GS can be related to the location of the chemical potential of the leads. 
For instance if the chemical potential is located between the GS and the $\gamma$GS, the 
GS($\gamma$GS) is located below(above) the Fermi function of the leads, respectively.
In this case, the current must flow from the left lead to the right lead via GS because it is below the Fermi function
and the opposite direction of flow may occur for the $\gamma$GS as it is above the Fermi function.

\begin{figure}[htb]
  \includegraphics[width=0.5\textwidth]{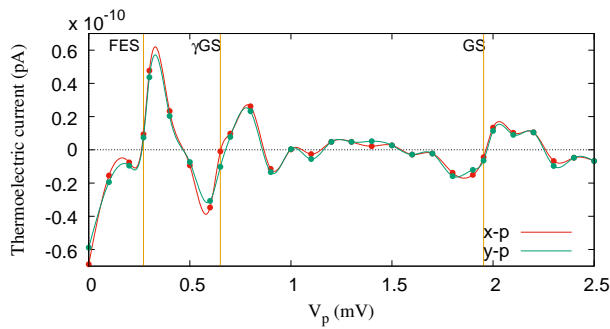}
 \caption{Thermoelectric current ($I_{\rm L}$) for the QD-system coupled to the photon field 
               with $x$- (red color) and $y$-polarized (green color) photon field.  
               The photon energy is $\hbar \omega_{\gamma} = 1.31$~meV (off-resonance regime),
               and $g_{\gamma} = 0.05$~meV, $\bar{n}_\textrm{R} = 1$.
               The temperature of the left (right) lead is fixed at ${\rm T_L} = 1.16$~K (${\rm T_R} = 0.58$~K) 
               implying thermal energy $0.1$~meV ($0.05$~meV), respectively.
               The chemical potential of the leads are fixed at $\mu_{\rm L} = \mu_{\rm R} = 1.2$~meV.
               The golden vertical lines show the resonance condition for the GS at ${\rm V_{p}} = 1.95$~mV, 
               the $\gamma$GS at ${\rm V_{p}} = 0.65$~mV, and the FES at ${\rm V_{p}} = 0.271$~meV, respectively.
               The weak external magnetic field is $B = 0.1~{\rm T}$, and $\hbar \Omega_0 = 2.0~{\rm meV}$.}
\label{fig06}
\end{figure}

The same explanation can be applied to the transport mechanism through the FES for the range $V_{\rm p} = [0 - 0.5]$~mV, but
instead of $\gamma$GS the one-photon replica of the first-excited state, $\gamma$FES, contributes to the transport here. 
Figure \ref{fig06} demonstrates the 
thermoelectric current versus the gate voltage for both $x$- (red color) and $y$-polarized photon field (green color).
It seems that the photon polarization for the off-resonance regime does not play an important role in the transport. 
The reason is that the location of the photon replica states in the MB-energy spectrum 
is not sufficiently changed by tuning the photon polarization from  the $x$- to the $y$-direction. 
Therefore, the contribution of the $\gamma$GS to the transport is almost the same for both polarization. 

The total electron occupation in the central system as displayed in Fig.\ \ref{fig04}(c) is similar to the results in
Fig.\ \ref{fig03}(c) except for the contribution around the $\gamma$GS peak, but the partial occupation shows strong
influences of the photon field. Around ${\rm V_{p}} = 1.8$~mV the GS is occupied as before, but now around 1/4 of the 
charge resides in the FES. A total change takes place for ${\rm V_{p}}$ in the range $0.0-1.0$~mV. There, now 60-70\% of the charge is
in the GS and the rest in the FES and the $\gamma$GS. If the time evolution is analyzed all the charge enters the 
central system through the FES and the $\gamma$GS, but the GS mainly gets occupied through slower radiative processes
made possible by the photon field.

\begin{figure}[htb]
  \includegraphics[width=0.5\textwidth]{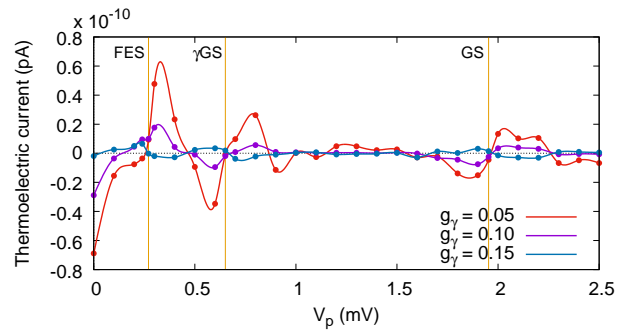}
 \caption{Thermoelectric current ($I_{\rm L}$) versus gate voltage for the QD-system coupled to the 
               photon cavity with electron-photon coupling strength $g_{\gamma} = 0.05$ (red color), $0.10$ (magenta color), and 
               $0.15$~meV (Navy blue color).
               The photon energy is $\hbar \omega_{\gamma} = 1.31$~meV (off-resonance regime),
               $\bar{n}_\textrm{R} = 1$, and the photon field is polarized in the $x$-direction.
               The temperature of the left (right) lead is fixed at ${\rm T_L} = 1.16$~K (${\rm T_R} = 0.58$~K) 
               implying thermal energy $0.1$~meV ($0.05$~meV), respectively.
               The chemical potential of the leads are fixed at $\mu_{\rm L} = \mu_{\rm R} = 1.2$~meV.
               The golden vertical lines show the resonance condition for the GS at ${\rm V_{p}} = 1.95$~mV, 
               the $\gamma$GS at ${\rm V_{p}} = 0.65$~mV, and the FES at ${\rm V_{p}} = 0.271$~meV, respectively.
               The weak external magnetic field is $B = 0.1~{\rm T}$, and $\hbar \Omega_0 = 2.0~{\rm meV}$.}
\label{fig07}
\end{figure}

To see further the influences of the photon field on thermal transport, we display the thermoelectric current for different 
electron-photon coupling strength, $g_{\gamma}$, in \fig{fig07} where $\bar{n}_\textrm{R} = 1$ and the photon field is polarized
in the $x$-direction. 
By increasing the electron-photon coupling strength, the thermoelectric current is suppressed
and a nearly zero current is recorded at $g_{\gamma} = 0.15$~meV over the same interval of voltage as before.
This is happens because the contributed ratio of the GS and the $\gamma$GS to the transport are almost equal
at a higher electron-photon coupling strength. As a result, the current is vanishing, and 
a plateau of nearly zero values is obtained.

\begin{figure}[htb]
   \includegraphics[width=0.23\textwidth]{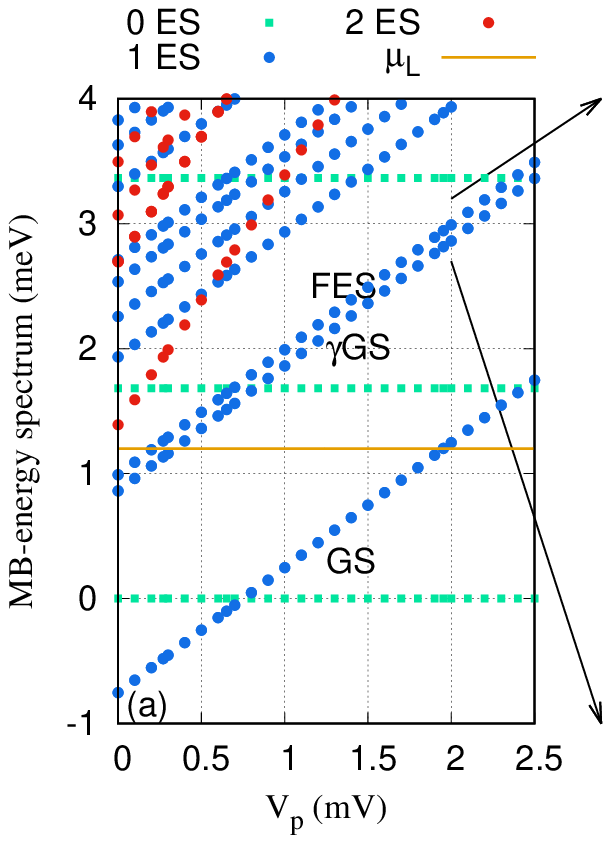}
  \includegraphics[width=0.23\textwidth]{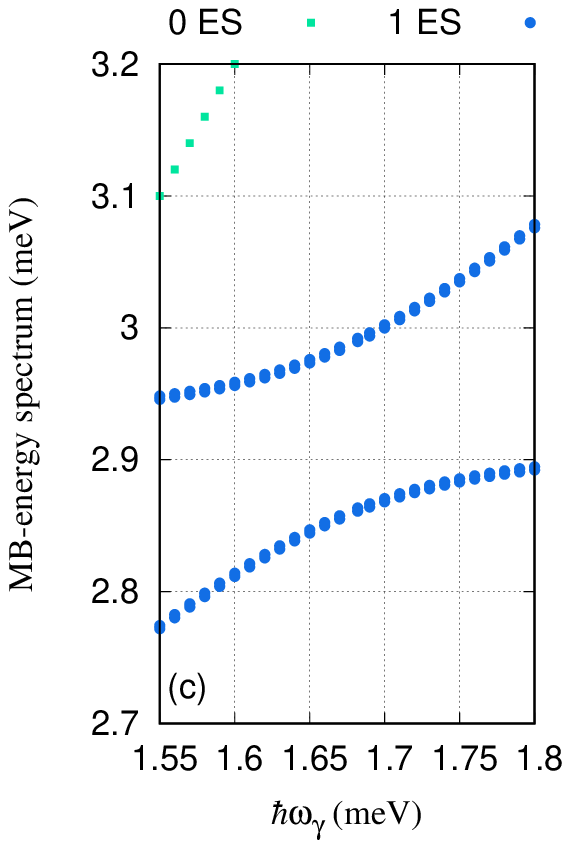}\\
     \includegraphics[width=0.23\textwidth]{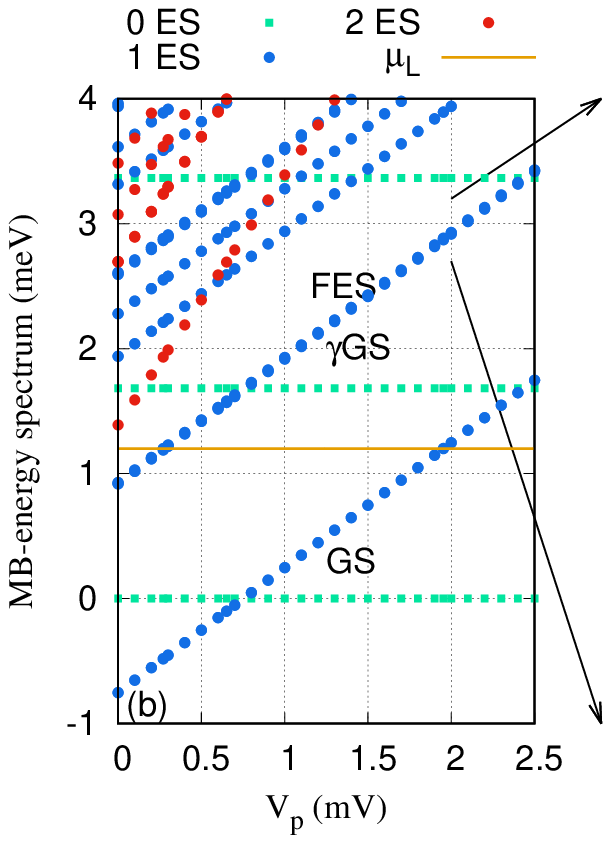}
  \includegraphics[width=0.23\textwidth]{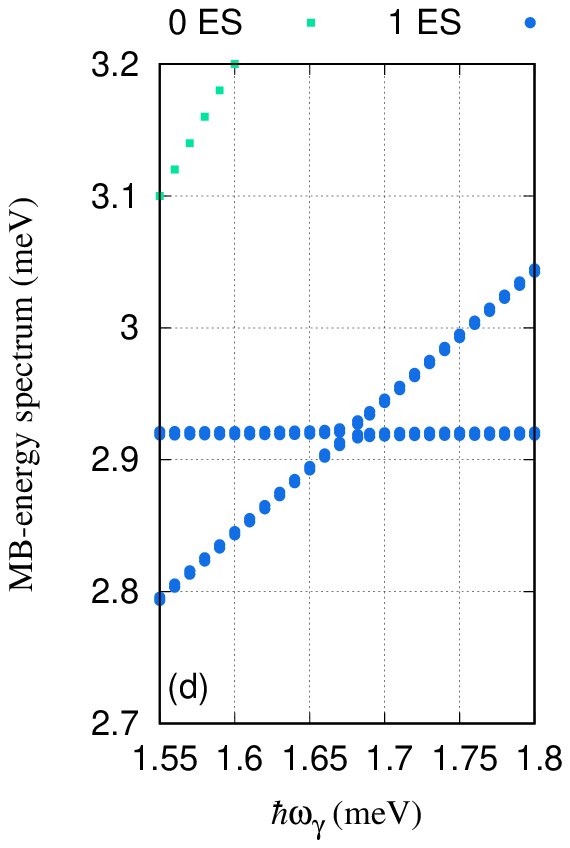}\\
 \caption{MB-Energy spectrum ($\rm E_{\rm \mu}$) of the quantum dot system versus the gate voltage ($\rm V_p$), 
         for the $x$- (a) and $y$-polarized (b) photon field. 
         The MB-energy spectrum of the $\gamma$GS and FES versus photon energy is plotted 
         for $x$- (c) and $y$-polarized (c) photon field. 
         The 0ES (green squares) are zero-electron states, 1ES (blue circles) are one-electron states, 
         and 2ES (red circles) are two-electron states.
         The golden line is the chemical potential of the leads where $\mu_L = \mu_R = \mu = 1.2~{\rm meV}$.
         GS indicates the one-electron ground-state, $\gamma$GS is the one-photon replica of the one-electron ground-state, and 
         FES is the one-electron first-excited state.
         The photon energy $\hbar \omega_{\gamma} = 1.68$~meV, and $g_{\gamma} = 0.05$~meV.
         The magnetic field is $B = 0.1~{\rm T}$, and $\hbar \Omega_0 = 2.0~{\rm meV}$.}
\label{fig08}
\end{figure}

We now investigate the resonant regime, when the photon energy is approximately equal to the 
energy spacing between the GS and the FES of the QD-system,
$\hbar \omega_{\gamma} \simeq E_{\rm FES} - E_{\rm GS}$. 
The photon energy is considered to be $\hbar \omega_{\gamma} =1.68$~meV,  
the electron-photon coupling strength is $g_{\gamma} = 0.05$~meV, and 
the mean photon number is $\bar{n}_\textrm{R} = 1$. 
The MB-energy spectrum as a function of gate voltage is shown in \fig{fig08} 
for the $x$- (a) and $y$-polarization (b) of the photon field. The Rabi-splitting  
between the $\gamma$GS and FES emerges and is larger for the $x$-polarized photon field. 
To confirm this we display the MB-energy spectrum of these two state as a function of the photon energy
for $x$- (\fig{fig08}c) and $y$-polarization (\fig{fig08}d). 
The anti-crossings at the photon energy $\hbar \omega_{\gamma} = 1.68$~meV indicates a Rabi-splitting 
and it is quite small for the $y$-polarized photon field.

\begin{figure}[htb]
  \includegraphics[width=0.5\textwidth]{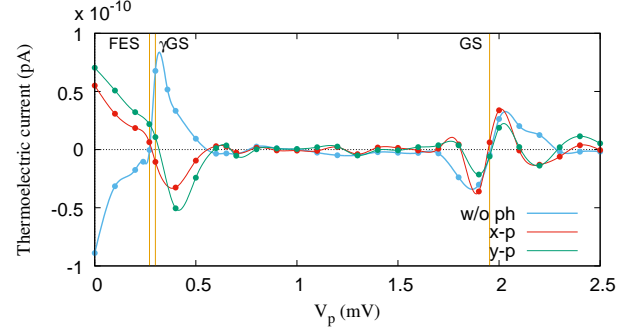}
 \caption{Thermoelectric current ($I_{\rm L}$) for the QD-system without (w/o ph) (blue color) 
               and with (w ph) $x$- (red color) and $y$-polarized (green color) photon field.  
               The photon energy is $\hbar \omega_{\gamma} = 1.68$~meV, $g_{\gamma} = 0.05$~meV, and $\bar{n}_\textrm{R} = 1$.
               The temperature of the left (right) lead is fixed at ${\rm T_L} = 1.16$~K (${\rm T_R} = 0.58$~K) 
               implying thermal energy $0.1$~meV ($0.05$~meV), respectively.
               The chemical potential of the leads are fixed at $\mu_{\rm L} = \mu_{\rm R} = 1.2$~meV.
               The golden vertical lines show the resonance condition for the GS at ${\rm V_{p}} = 1.95$~mV, 
               and the Rabi-splitting states between the $\gamma$GS and the FES at ${\rm V_{p}} = 0.271$~meV, respectively.
               The weak external magnetic field is $B = 0.1~{\rm T}$, and $\hbar \Omega_0 = 2.0~{\rm meV}$.}
\label{fig09}
\end{figure}

The thermoelectric current for the on-resonant regime is shown in \fig{fig09}. 
We find that the thermoelectric current through the GS is almost unchanged for the both polarizations,
but the characteristics of the thermoelectric current of the FES, that is in resonance with the $\gamma$GS, 
is drastically modified.
The effect of the resonant photon field is to invert the thermoelectric current from `positive' to `negative' values or vice versa
around the FES at $V_{\rm p} = 0.271$~mV. The more $\gamma$GS like state at $V_{\rm p} \simeq 0.279$~mV contributes 
to the electron transport with the more FES like state leading to the current flip from `positive' to `negative' values.
In addition, the first photon replica of the first-excited state ($\gamma$FES) becomes active in the transport here.
It should be noted that the current inversion is larger for the smaller Rabi-splitting in the $y$-polarized photon field. 
It indicates that the photon replica states have a major contribution to the transport, the resonance condition activates
higher lying states in the spectrum in the transport.

\section{Conclusion}\label{Sec:Conclusion}

The characteristics of thermoelectric transport 
through a quantum dot embedded in a short quantum wire interacting with either off-
or on-resonant cavity-photon field have been investigated in a steady-state regime.
The QD-system is considered to be connected to two electron reservoirs
with different temperatures. The temperature gradient can accelerate electrons
from either lead to the QD-system generating a thermoelectric current.

If a linearly polarized photon field is applied, the properties of the thermoelectric current are drastically changed. 
In the off-resonant regime, when the photon energy is smaller than the two lowest energy states of the 
QD-system, additional current peak is observed which is only caused by photon-induced or photon assisted transport.
In addition to that a plateau in the thermoelectric current is formed at a high electron-photon coupling strength.
In the resonant regime, the effects of a Rabi-splitting in the energy spectrum appears 
leading to an inversion of thermoelectric current. The inversion is more drastic for the smaller Rabi-splitting
observed for a $y$-polarized photon field.

Even though we set out to investigate the thermoelectric transport in the challenging regime of Coulomb blockades
interrupted by narrow resonant peaks as the plunger gate voltage is changed, we observe that our approach indicates 
a mechanism that could be used in a thermoelectric inversion device.

\begin{acknowledgments}

This work was financially supported by the Research Fund of the University of Iceland,
the Icelandic Research Fund, grant no.\ 163082-051, 
and the Icelandic Infrastructure Fund. 
The computations were performed on resources provided by the Icelandic 
High Performance Computing Center at the University of Iceland.
NRA acknowledges support from University of Sulaimani and 
Komar University of Science and Technology.
CST acknowledges support from Ministry of Science and
Technology of Taiwan under grant No.\ 106-2112-M-239-001-MY3.

\end{acknowledgments}

%-----------------------------------------------------------------
%
%\bibliographystyle{unsrt}
\bibliographystyle{apsrev4-1}
%\bibliography{Ref.bib}
%
%
%----------------------------------------------------------------------------------------
%
\end{document}